\documentclass[aps,prd,groupedaddress,onecolumn,showpacs,10pt]{revtex4-1}
\usepackage{graphicx,color}
\usepackage{bm}
\usepackage{amssymb,amsmath,amsfonts, mathrsfs}

\newcommand{\be}{\begin{equation}}
\newcommand{\ee}{\end{equation}}
\newcommand{\ba}{\begin{array}}
\newcommand{\ea}{\end{array}}
\newcommand{\bqa}{\begin{eqnarray}}
\newcommand{\eqa}{\end{eqnarray}}

\begin{document}


\title{Comprehending Isospin breaking effects of  $X(3872)$  in a Friedrichs-model-like scheme}


\author{Zhi-Yong Zhou}
\email[]{zhouzhy@seu.edu.cn}
\affiliation{School of Physics, Southeast University, Nanjing 211189,
P.~R.~China}
\author{Zhiguang Xiao}
\email[]{xiaozg@ustc.edu.cn}
\affiliation{Interdisciplinary Center for Theoretical Study, University of Science
and Technology of China, Hefei, Anhui 230026, China}


\date{\today}

\begin{abstract}
Recently, we have shown that the $X(3872)$ state can be naturally generated as a
bound state by incorporating the hadron interactions
into the Godfrey-Isgur quark model using the Friedrichs-like model combined
with the QPC model, in which the  wave function for the $X(3872)$ as a combination of the bare $c\bar
c$ state and the continuum states can also be obtained. Under this scheme,
we now 
 investigate the isospin breaking effect of $X(3872)$ in its decays to
$J/\psi\pi^+\pi^-$ and $J/\psi\pi^+\pi^-\pi^0$. By Considering its dominant
continuum parts coupling to $J/\psi\rho$ and $J/\psi\omega$ through
the quark rearrangement process, one could obtain the reasonable ratio
of $\mathcal{B}(X(3872)\rightarrow
J/\psi\pi^+\pi^-\pi^0)/\mathcal{B}(X(3872)\rightarrow
J/\psi\pi^+\pi^-)\simeq (0.58\sim 0.92)$.
It is also shown that the $\bar D D^*$ invariant mass distributions in the $B\rightarrow \bar D D^* K$ decays could be understood qualitatively at the same time. This scheme may provide more insight to understand the enigmatic nature of the $X(3872)$ state.
\end{abstract}


\maketitle
\section{Introduction}
The enigmatic $X(3872)$ state has been highlighted for more than one
decade, after it was first discovered by Belle~\cite{Choi:2003ue} and
confirmed by CDF, D0, and
\textsl{BABAR}~\cite{Acosta:2003zx,Abazov:2004kp,Aubert:2004ns}.
The mass of the $X(3872)$ is at $3871.69\pm 0.17$
MeV~\cite{Olive:2016xmw}, almost degenerate with the $D^{*0}\bar D^0$  threshold, which is
the most intriguing feature. Its width is also very narrow, with an
upper limit $1.2\ \mathrm{MeV}$.  Its
quantum number is determined to be $J^{PC}=1^{++}$ by
LHCb~\cite{Aaij:2013zoa}, which is consistent with its radiative
decay~\cite{Bhardwaj:2011dj,Aubert:2006aj,Bhardwaj:2011dj} and
multi-pion transitions~\cite{Abulencia:2005zc,delAmoSanchez:2010jr}.
The negative result in searching for its charged partner in $B$
decays~\cite{Aubert:2004zr} implies that it should be an isoscalar
state. Nevertheless, the dominant $\rho$ contribution of dipion mass
spectrum in $X(3872)\rightarrow
J/\psi\pi^+\pi^-$~\cite{Acosta:2003zx}  suggests a large isospin
breaking effect. Compared with its decay to $J/\psi\pi^+\pi^-\pi^0$
through the I=0 $\omega$ resonance, the ratio is measured to be
\bqa \frac{
\Gamma(X(3872)\rightarrow J/\psi\pi^+\pi^-\pi^0)}{
\Gamma(X(3872)\rightarrow J/\psi\pi^+\pi^-)}=1.0\pm 0.4\pm 0.3
\eqa
by Belle~\cite{Abe:2005ix} and
\bqa
\frac{
\Gamma(X(3872)\rightarrow J/\psi\pi^+\pi^-\pi^0)}{
\Gamma(X(3872)\rightarrow J/\psi\pi^+\pi^-)}=0.8\pm 0.3
\eqa
 by
\textsl{BABAR}~\cite{delAmoSanchez:2010jr}.
These characteristics and
other properties as discussed in the literatures~(see
refs.~\cite{Guo:2017jvc,Chen:2016qju,Esposito:2016noz,Lebed:2016hpi}
for example) suggest an exotic nature of the $X(3872)$ .

The proximity of the $X(3872)$ to the $D^{*0}\bar D^0$ threshold leads to
a direct interpretation of it as a hadronic molecular state. In fact,
T\"ornqvist predicted a similar $D\bar D^*$ bound state around
$3870$MeV about ten years
before the discovery of the $X(3872)$\cite{Tornqvist:1993ng}, using a
pion exchange model similar to the discussion of the Deuteron in the
$pn$ system, and
named the state as a ``deuson''. However, this picture may not
explain the production cross section of $X(3872)$ in $p\bar p$
annihilation~\cite{Bignamini:2009sk} and
also can not explain its large decay rate to $\gamma \psi'$~\cite{Swanson:2003tb,Dong:2009uf}.
This meson-exchange model is also
extended to include other intermediate hadronic states like $\sigma$,
$\rho$ and $\omega$ in
Ref.~\cite{Liu:2008fh,Liu:2008tn,Li:2009zu}.
The tetraquark model is also
introduced to explain its existence and the other exotic
states\cite{Maiani:2004vq}. However, this explanation also meets the large $\gamma\psi'$ decay
rate problem. This bound state can also be reproduced by other models,
such as
the effective Lagrangian approach~\cite{AlFiky:2005jd,Fleming:2007rp},
 chiral unitary
calculation~\cite{Wang:2013kva}, screened potential
model~\cite{Li:2009zu} and coupled-channel model~\cite{Zhou:2013ada}.
It was also pointed out in~\cite{Braaten:2007dw, Coito:2012vf} that $X(3872)$ could
be a mixture of a $c\bar c$ and the molecule-like $D\bar D^*$
components which may provide both solution to the production and the
radiative decay problem.
In our recent work~\cite{Zhou:2017dwj}, we discussed the first exited $P$-wave charmonium
states using a Friedrichs-like model
combined with the quark pair creation~(QPC) model~\cite{Micu:1968mk,Blundell:1995ev} to incorporate the hadron interaction effect
into the Godfrey-Isgur~(GI) quark model.   We found that
 in the $2^3P_1$ channel the $X(3872)$ could be naturally generated as a bound state just below
the $D^{0}\bar{D}^{0*}$ threshold in this framework, while
another resonance pole at around  $3.934\pm 0.040 i$GeV on the Riemann
sheet attached to the physical region above the $\bar D D^*$ threshold is found to be
generated from the bare $\chi_{c1}(2P)$, which might be related to the
observed $X(3940)$ state in the experiment.
Compared to the
above ``deuson" picture, in this scheme, similar to
Ref.~\cite{Braaten:2007dw, Coito:2012vf}, the $X(3872)$ is dynamically generated
by the coupling of the bare ${2}^3P_1$ $c\bar c$ state with the continuum,
not from the pion exchange. The wave function of
$X(3872)$ can also be explicitly written down as a linear combination of $c\bar
c$ and the continuum states.  With this information, more properties of
the $X(3872)$ can be studied. In this paper, we will mainly focus on
the isospin breaking effect based on this result.

The ratio ${ \Gamma(X(3872)\rightarrow J/\psi\pi^+\pi^-\pi^0)}/{
\Gamma(X(3872)\rightarrow J/\psi\pi^+\pi^-)}$ seems to imply a
significant isospin breaking effect, but it is first pointed out by
Suzuki that it might be misleading~\cite{Suzuki:2005ha}. Because the central mass value of $J/\psi\omega$ is about
7 MeV higher than the $X(3872)$ mass, while the central value of
$J/\psi\rho$ energy is lower
than the $X(3872)$ mass, the $\pi^+\pi^-\pi^0$ in the
$J/\psi\pi^+\pi^-\pi^0$ final state comes from the far tail of the
$\omega$ resonance and the kinematical phase space of the isospin-conserved process
is highly suppressed. Considering this
kinematical constraint, the production amplitude ratio
$\frac{A(X(3872)\rightarrow J/\psi\rho)}{A(X(3872\rightarrow
J/\psi\omega))}$ is estimated to be $0.27\pm 0.02$ to produce a comparable result with
the experiment. Meng and Chao addressed this problem by considering
the rescattering effect in an effective lagrangian method~\cite{Meng:2007cx}. They treated
the $X(3872)$ as an elementary field and introduced the effective
interactions among the $X$ state, the pseudoscalar, and the vector
mesons. By calculating the imaginary and real part of the rescattering
amplitudes, they found a consistent value of the ratio in their
parameter space.  Li and Zhu~\cite{Li:2012cs} investigated the
probability in the framework of one-meson-exchange with considering
the S-D wave mixing and they obtained the ratio of these two modes to
be 0.42. Gammermann and Oset~\cite{Gamermann:2009fv} used the
on-mass-shell Lippmann-Shwinger equation method and found a branch
ratio of 1.4.

In the present paper, as stated above, we will discuss the
isospin-breaking effect from the starting point of our previous
paper~\cite{Zhou:2017dwj}. The Friedrichs-like model is exactly
solvable and the interactions between the bare $c\bar c$ and the
Okubo-Zweig-Iizuka~(OZI) allowed continuum states are approximated by
the QPC model using the wave
functions for bare states from the GI model. This just provides a general scheme to make corrections
to the well-accepted GI model by
including the hadron interactions. The experimentally observed
first excited $P$-wave charmonium states can be reasonably produced
simultaneously.
The wave functions for these states in terms of the bare $c\bar c$ and
continuum states can also be explicitly written down.
Once the wave function of the $X(3872)$
is explicitly given in this scheme, we are able to insert an
interaction Hamiltonian between the $X(3872)$ wave function and the
final states $J/\psi\omega$ or $J/\psi\rho$ and calculate the related
transition amplitudes and the branching fraction. If the OZI suppressed
couplings are omitted, the $c\bar c$ contribution would be neglected
and only the continuum components will contribute to the amplitude. The
couplings between the final states and the continuum parts could then be
described by the quark rearrangement model developed by Barnes and
Swanson~(BS model)~\cite{Barnes:1991em}. The merit of this choice is that the BS model
considers the spin-spin hyperfine, color Coulomb, and linear confinement interactions among
the quarks between different mesons, which respects the same spirit as
the GI model and  no new parameter needs to be  introduced. By a
standard derivation of the representation of transition amplitude and
numerical calculations, our final result of $\frac{
\Gamma(X(3872)\rightarrow J/\psi\pi^+\pi^-\pi^0)}{
\Gamma(X(3872)\rightarrow J/\psi\pi^+\pi^-)}$ turns out to be in good
agreement with the experiment value.  Our method may have some
similarity  with the
widely used coupled-channel formalism such as~\cite{Eichten:1978tg,
Kalashnikova:2005ui,Ortega:2009hj,Takizawa:2012hy}.
However, we are using the well-accepted GI's wave function in
the QPC model and the quark rearrangement model in our calculation, which
provides more solid theoretical basis and makes our result more
convincing.

This paper is organized as follows: The theoretical foundations are
introduced in Section II, which include the Friedrichs model, its
extended scheme, and the quark rearrangement model by Barnes and
Swanson. The    numerical calculations of the isospin breaking
effect are given in
Section III. Section IV is the final conclusion and discussions.

\section{Theoretical model}

In this section, we will briefly introduce the theoretical basis of our
discussion which includes the main results of the Friedrichs model and
its extended version, the quark rearrangement model by Barnes and
Swanson, and the derivations of the transition amplitudes of $X(3872)$
to $J/\psi\omega$ or $J/\psi\rho$.

\subsection{The Friedrichs model}
In 1948, Friedrichs proposed an exactly solvable model to understand
an
unstable state~\cite{Friedrichs:1948}. The simplest form of Friedrichs
model includes a free Hamiltonian $H_0$  and an interaction part $V$.
The free Hamiltonian $H_0$ has a bare continuous spectrum
$[E_{th},\infty)$, and a discrete eigenvalue $E_0$ imbedded in this
continuous spectrum~($E_0 >E_{th}$). The interaction part describes
the coupling between the bare continuous state and discrete  states of
$H_0$ so that the discrete state is dissolved in the continuous state
and a resonance is produced.

In the energy representation, the free Hamiltonian can be expressed as
\begin{align}
H_0=E_0|0\rangle\langle
0|+\int_{E_{th}}^\infty E|E\rangle\langle E|\mathrm{d}E,
\end{align}
where $|0\rangle$ denotes the bare discrete state and
$|E\rangle$ denotes the bare continuum state.
The normalization conditions for the bare states are
\bqa\label{normal}
\langle 0|0\rangle=1, \langle E|E'\rangle=\delta(E-E'),\langle 0|E\rangle=\langle E|0\rangle=0.
\eqa
The interaction part serves to couple the bare discrete state and the bare continuous state as
\bqa
V=\lambda\int_{E_{th}}^\infty \big[f(E)|E\rangle\langle 0
|+f^*(E)|0\rangle\langle E |\big]\mathrm{d}E,
\eqa
where $f(E)$ function denotes the coupling function and $\lambda$  the coupling
strength.  This eigenvalue problem  $H\Psi(x)=x\Psi(x)$, with the full Hamiltonian $H=H_0+V$, can be exactly
solved and the final continuum state is
\begin{eqnarray}
|\Psi_\pm(x)\rangle&=&|x\rangle+\lambda\frac{f^*(x)}{\eta^\pm(x)}\Big[|0\rangle+\lambda\int_{E_{th}}^\infty\frac{f(E)}{x-E\pm
i \epsilon}|E\rangle\mathrm{d}E\Big],
\label{eq:continuum-state}
\end{eqnarray}
in which the resolvent $\eta^{\pm}(x)$ is defined as
\begin{align}
\eta^{\pm}(x)=x-E_0-\lambda^2\int_{E_{th}}^\infty\frac{|f(E)|^2}{x-E\pm
i \epsilon}\mathrm{d}E\,.\label{eq:eta-pm}
\end{align}
The normalization is $\langle\Psi_\pm
(E)|\Psi_\pm(E')\rangle=\delta(E-E')$ and the subscript $\pm$ denotes
the in-states~($+$) and out-states~($-$), respectively. The scattering $S$-matrix can also be obtained as
\bqa\label{scatteringSmatrix}
S(E,E')=\delta(E-E')\Big(1-2\pi i \frac{\lambda f(E)f^*(E)}{\eta^+(E)}\Big).
\eqa
The $\eta^\pm$ functions can be analytically continued to the complex
$z$-plane to be one complex function $\eta(z)$  for $z\in \mathbb{C}$
with its boundary values $\eta(x\pm i\epsilon)=\eta^\pm(x)$ on the
real axis. Since there is only one threshold for the continuum, there
is only one cut for this function, and the $\eta(z)$ function is
defined on a two-sheeted Riemann surface. The zero points for the
$\eta(z)=0$ will be the poles for the $S$-matrix.  The zero points of
$\eta(z)$ or the poles of the $S$-matrix will represent the
generalized complex-valued eigenstates~(called Gamow state) for the
full Hamiltonian, which satisfy $H|z\rangle=z|z\rangle$ with $z\in
\mathbb{C}$ as described in the Rigged-Hilbert-Space~(RHS) formulation
of the quantum mechanics developed by Bohm and
Gadella~\cite{Bohm:1989,Civitarese200441}. By summing all the
perturbation series, Prigogine and his collaborators also obtained a
similar mathematical description~\cite{Prigogine:1991}.  In general,
the different kinds of the generalized eigenstates  are summarized in
the following:

\begin{enumerate}
  \item Bound state:

  The solution of $\eta(z)=0$ on the first-sheet real axis below
$E_{th}$ represents a bound state. Its wave function is written down as
  \bqa
|z_{B}\rangle=N_{B}
\Big(|0\rangle+\lambda\int_{E_{th}}^\infty\frac{f(E)}{z_{B}-E}|E\rangle\mathrm{d}E\Big),
\label{eq:Bound}
\eqa
This solution has a finite norm and can be normalized as $\langle
z_B|z_B\rangle=1$ where
\bqa N_B=(\eta'(z_B))^{-1/2}=\Big(1+\lambda^2\int \mathrm{d}E \frac{|f(E)|^2}{(z_B-E)^2}\Big)^{-1/2}.
\eqa
Then, it is straightforward  to define the so-called ``elementariness" $Z$ and ``compositeness" $X$ as
\bqa\label{eq:compositeness}
Z=\frac{1}{N_B^2},\,\,\,\, X=\frac{\lambda^2}{N_B^2}\int \mathrm{d}E \frac{|f(E)|^2}{(z_B-E)^2}
\eqa
similar to the ones in the Weinberg's pioneering work in studying the
compositeness of deuteron~\cite{Weinberg:1962hj}. The physical
meanings of the ``elementariness" and ``compositeness"  are the
probabilities of finding the bare discrete and the bare continuum
states in the bound state respectively.

\item Virtual state:

  The solution lying on the real axis of the second Riemann sheet
below the threshold represents a virtual state. Its wave function is
expressed as
  \bqa
  |z^\pm_{V}\rangle=N_{V}
\Big(|0\rangle+\lambda\int_{E_{th}}^\infty\frac{f(E)}{z_{V}^\pm-E}|E\rangle\mathrm{d}E\Big),
  \label{eq:Virtual}
  \eqa
where the superscript $\pm$ denotes the two kinds of integration
contours which are continued from the first sheet to the second sheet
to enclose the virtual state pole from the upper side (+) or the lower
side (-) of the first sheet cut~\cite{Xiao:2016dsx}. Unlike the bound
state, a virtual state does not have a well-defined norm as the usual
state in the Hilbert space. Thus the compositeness and the
elementariness for a virtual state can not be mathematically
rigorously defined. However, we can define a normalization such that
$\langle z_V^-|z_V^+\rangle=1$, by choosing
\bqa
N_V=(\eta'(z_V))^{-1/2}=\Big(1+\lambda^2\int \mathrm{d}E \frac{|f(E)|^2}{(z_V-E)^2}\Big)^{-1/2}.
\eqa
One typical example  is the virtual state in the singlet neutron-proton scattering,
which serves to contribute to the unusually large scattering
length.

\item Resonant state:

The real analyticity of the $\eta(z)$ function requires the solutions
of $\eta(z)=0$  on the second complex energy plane to appear as a
conjugate pair. This pair of poles of $S$-matrix represents a
resonance since it will be unstable for its finite imaginary part of
the energy eigenvalue. The pole position on the lower-half energy
plane of the second Riemann sheet is related to the mass $M$ and width
$\Gamma$ as $z_R=M-i\frac{\Gamma}{2}$. The wave functions for them is
written as
\begin{eqnarray}
|z_R\rangle=N_R\Big(|0\rangle+\lambda\int_{E_{th}}^\infty\mathrm{d}E\frac{f(E)}{[z_R-E]_+}|E\rangle\Big),\nonumber\\
|z_R^*\rangle=N_R^*\Big(|0\rangle+\lambda\int_{E_{th}}^\infty\mathrm{d}E\frac{f^*(E)}{[z_R^*-E]_-}|E\rangle\Big),
\label{eq:Gamow-state-right}
\end{eqnarray}
where $z_R$ is on the lower half plane and $z_R^*$ is its complex
conjugate. The $[\cdots]_\pm$ means the analytical continuations of
the integrations from upper or lower edge of the first sheet
cut~\cite{Xiao:2016dsx}. Similar to the virtual state, the wave
function of a resonant state can not be normalized as usual, and
thus is
not the normal state vector in the Hilbert space. However we can
choose
\bqa
N_R=(\eta '^+(z_R))^{-1/2}=\Big(1+\lambda^2\int
\mathrm{d}E \frac{|f(E)|^2}{[(z_R-E)_+]^2}\Big)^{-1/2}
\eqa
to normalize the state as $\langle z_R^*|z_R\rangle=1$ and similarly
for the $|z_R^*\rangle$. The normalization factor $N_R$ will be
complex in general and one might define ``elementariness" and
``compositeness" parameters as in Ref.~\cite{Xiao:2016mon}, but their
physical meanings are not clear because they are complex numbers too.
However, some other physical approximate definitions proposed in
Ref.~\cite{Sekihara:2014kya,Guo:2015daa} might be used to describe
these quantities.
\end{enumerate}
It is worth emphasizing that these discrete spectrum solutions may or
may not be generated from the bare discrete state of the free
Hamiltonian. If the state is generated from the bare discrete state,
it will move back to the bare state as one decreases the coupling
strength to zero. Besides, the discrete state could also be dynamically
generated from the singularities of the form factor. In this occasion,
if the coupling strength is tuned to be zero, the positions of this
kind of state may move towards the singularity of the form factor on
the unphysical sheets~\cite{Xiao:2016dsx,Xiao:2016wbs}.  It is also
worth mentioning that the Friedrichs model has several variants such
as the Fano model~\cite{Fano:1961zz}, the Lee model~\cite{Lee:1954iq},
the Anderson model~\cite{Anderson:1961} in other physics areas.

\subsection{The extended Friedrichs model and the QPC model}

In the original Friedrichs model, the states are labeled only by the
energy quantum number which seems to be unrelated to the states in the
three dimensional space. In fact, after partial-wave decomposition of
the three dimensional states, the similar model in terms of the
angular momentum eigenstate is reduced to a Friedrichs-like
model~\cite{Xiao:2016mon}. Consider the coupling of a bare discrete
state $|0,JM\rangle$ with a spin quantum number $J$ and the magnetic
quantum number $M$ and the bare continuum momentum eigenstate
$|\vec{p},S,S_z\rangle$ with a total spin $S$, $z$-component $S_z$,
and $\pm \vec p$ the c.m. momentum for the two particles composing the
continuum state for example. The discrete state denoted by
the total Hamiltonian for fixed $JM$ can be
recast into
\begin{align}
H=& M_0 |0\rangle\langle0|
+\sum_{L}
\int \mathrm d E\, E |E,L\rangle \langle E,L|
+\sum_L\int \mathrm{d}E  f_L(E)|0\rangle\langle
E,L|+h.c.
\end{align}
in which $|0\rangle=|0;JM\rangle$, $|E,L\rangle=\sqrt{\mu
p}|p;JM;LS\rangle$ with $\mu$ the reduced mass of the two-particle state in its $c.m.$ frame and $f(E)$ the coupling form factor, as derived in Ref.~\cite{Xiao:2016mon}.
This is just similar to the original Friedrichs model but with more
continuum and the similar exact solution can be obtained.

We can make more generalization by adding more discrete states and
more continuum states, and the interactions between continuum states
can also be introduced. The most general Hamiltonian with $D$ discrete
states, $|i\rangle$  ($i=1,\dots,D$), and $C$ continuum states,
$|E_j,j \rangle$ ($j=1,\dots,C$), can be expressed as
\begin{align}
H=&\sum_{i=1}^D M_i |0;i\rangle\langle0;i|+\sum_{i=1}^C\int_{M_{i,th}}^\infty \mathrm{d}E_i E_i
|E_i;i\rangle\langle E_i;i|
\nonumber\\&+\sum_{i_2,i_1}\int_{M_{i_1,th}}\mathrm d E'\,\int_{M_{i_2,th}}\mathrm
d E\,
 g_{i_2,i_1}(E',E)|
E';i_2\rangle \langle E;i_1| +h.c.
\nonumber\\&+\sum^D_{i=1}\sum_{j=1}^C\int_{M_{j,th}} \mathrm{d}E
f_{i,j}(E)|0;i\rangle\langle
E;j|+h.c.
\label{eq:generalFriedrichs}
\end{align}
where $f_{i,j}(E)$ is the form factor describing the interaction
between the $i$th discrete state and the $j$th continuum state, and
$g_{ij}$  describes the interaction between the $i$th continuum and
the $j$th continuum. For general interactions $g_{ij}$, the model is
not solvable, but if $g_{ij}(E',E)=v_{ij}f_i(E')f_j(E)$
and $f_{ij}=u_{ij}f_j(E)$, where $u_{ij}$ and $v_{ij}$ are
constant, the model can also be exactly solved. See Ref.
\cite{Xiao:2016mon} for details.

 In the present study, we only consider the case where only one
discrete states is coupled with several continuum states without the
interactions between the continuum states. The solutions differ from
the ones previously described by simply adding more similar continuum
integrals in Eqs.~(\ref{eq:eta-pm}), (\ref{eq:Bound}),
(\ref{eq:Virtual}), (\ref{eq:Gamow-state-right}).

We will restrict our study on the properties of mesons. The next
problem is to determine the interaction between one meson state
and the two meson continuum states, $i.e.$ the coupling form factor in the
Friedrichs model. A simple method is to use the QPC
model~\cite{Micu:1968mk,Blundell:1995ev} to describe the interaction
of one meson and two-meson continuum states.

In this model, the meson coupling $A\rightarrow BC$ can be defined as the
transition matrix element
\bqa\langle BC|T|A\rangle=\delta^3(\vec{P_f}-\vec{P_i})\mathcal M^{ABC}
\label{eq:MABC}
\eqa
where
the transition operator $T$  in the QPC model is defined as
\bqa
T=-3\gamma\sum_m\langle 1 m 1 -m|00\rangle\int \mathrm{d}^3\vec{p_3}\mathrm{d}^3\vec{p_4}\delta^3(\vec{p_3}+\vec{p_4})
\mathcal{Y}_1^m(\frac{\vec{p_3}-\vec{p_4}}{2})\chi_{1 -m}^{34}\phi_0^{34}\omega_0^{34}b_3^\dagger(\vec{p_3})d_4^\dagger(\vec{p_4}),
\eqa
describing a quark-antiquark pair generated by the $b^\dagger_3$ and
$d^\dagger_4$ creation operators from the vacuum.
$\phi_0^{34}=(u\bar u+d\bar d+s\bar s)/\sqrt{3}$ is the $SU(3)$ flavor
wave function for the quark-antiquark pair. $\chi^{34}_{1-m}$ and $\omega^{34}_0$
are the spin wave function and the color wave function, respectively.
$\mathcal Y_1^m$ is the solid Harmonic function. $\gamma$
parameterizes the production strength of the quark-antiquark pair from the vacuum.
The definition of the meson state here is different from the one in
Ref.~\cite{Hayne:1981zy} by omitting the factor $\sqrt{2E}$ to ensure
the correct normalizations in the Friedrichs-like model, as
\begin{align*}
|A(n, { }^{2S+1}L_{J,M})(\vec
P)\rangle=&\sum_{M_L,M_S}\langle LM_LSM_S
|JM\rangle\int \mathrm{d}^3p\,
\psi_{nLM_L}(\vec p)\,\chi^{12}_{SM_S}\,\phi^{12}\,\omega^{12}
\\&\times\Big|q_1\Big(\frac
{m_1}{m_1+m_2}\vec P+\vec p\Big)\bar q_2\Big(\frac {m_2}{m_1+m_2}\vec
P-\vec p\Big)\Big\rangle.
\end{align*}
 $\chi^{12}$, $\phi^{12}$ and $\omega^{12}$  are the spin wavefunction, flavor wave
function and the color wave function, respectively. $p_1$ ($p_2$) and
$m_1$ ($m_2$)
are the momentum and mass of the quark (anti-quark).   $\vec P=\vec
p_1+\vec p_2$ is the momentum of the  center of mass,
and $\vec p=\frac{m_2\vec p_1-m_1\vec p_2}{m_1+m_2}$ is the
relative momentum. $\psi_{nLM_L}$ is the wave function for the meson,
$n$ being the radial quantum number.

By the standard derivation one can obtain  the amplitude
$\mathcal M^{ABC}$ defined by Eq. (\ref{eq:MABC})
and the partial-wave amplitude $\mathcal M^{SL}(P(E))$
as in Ref.~\cite{Blundell:1995ev}. Then the form factor $f_{SL}$ which
describes the interaction between $|A\rangle $ and $|BC\rangle$  in the
Friedrichs model can be obtained  as
\begin{align}
f_{SL}(E)=\sqrt{\mu P(E)}\mathcal M^{SL}(P(E)),
\end{align}
where $P(E)=$$\sqrt{\frac{2M_BM_C(E-M_B-M_C)}{M_B+M_C}}$ is
the c.m. momentum,  $M_B$ and $M_C$ being the masses of meson $B$ and
$C$ respectively.

Now, the full Hamiltonian $H$ can be expressed as
\begin{align}\label{fullHamiltonian}
H=&E_0 |0\rangle\langle 0| +\sum_{n,S,L}
\int_{E_{th,n}}^\infty \mathrm dE\, E |E;n,SL\rangle \langle E;n,SL|
 \,\nonumber\\+&\sum_{n,S,L}\int_{E_{th,n}}^\infty \mathrm{d}E  f^n_{SL}(E)|0\rangle\langle
E;n,SL|+h.c.
\end{align}
where $E_0$ denotes the bare mass of the discrete state,  $n$ the
$n$-th continuum state, $E_{th,n}$ the energy threshold of $n$-th
continuum state, $S$ and $L$ the total spin and the angular momentum
of the continuum states, $f^n_{SL}$ the coupling functions between the
bare state and the $n$-th continuum state with particular $S,L$
quantum numbers. The eigenvalue problem of the full Hamiltonian in
Eq.~(\ref{fullHamiltonian}) is exactly solvable as mentioned above.
The final eigenlvalues of  bound states, virtual states or resonant
states could be obtained by solving the resolvent function $\eta(z)=0$
on the complex energy plane where
\bqa
\label{resolvent}
\eta(z)=z-E_0&-&\sum_n\int_{E_{th,n}}^\infty\frac{\sum_{S,L}|f_{SL}^n(E)|^2}{z-E
}\mathrm{d}E.
\eqa

In Ref.~\cite{Zhou:2017dwj}, the $X(3872)$ and other first excited
charmonium-like states could be generated based on the parameters and
the wave functions of the GI model.
$\gamma\sim 4.0$
 is chosen  such that the $X(3872)$ state
emerges as a bound state pole at  around $3871.4$MeV, just below the $D^0\bar{D}^{0*}$ threshold,
which is not generated from the bare discrete state. This value is
reasonable because the $\chi_{c2}(2P)$, the $X(3872)$ in both
channels match the experimental value simultaneously and a state
around $3933$MeV which is generated from $\chi_{c1}(2P)$ can also be
assigned to the experimental observed $X(3940)$.  These information
provides the starting point for discussing the isospin breaking
property in  the decays of the $X(3872)$ state.  In this setup, the
wave function of the $X(3872)$ is expressed explicitly as
\begin{align}
|X(3872)\rangle=&N_B\Big(|c\bar c\rangle+\int_{M_{00V}}^\infty \mathrm{d}E
\sum_{S,L}\frac{f^{00V}_{SL}(E)} {z_X-E}(|E;D^0\bar
D^{0*},SL\rangle+|E;D^{0*}\bar D^0,SL\rangle
)\nonumber
\\+&\int_{M_{+-V}}^\infty
\mathrm{d}E\sum_{S,L}\frac{f^{+-V}_{SL}(E)}
{z_X-E}(|E;D^+D^{-*},SL\rangle+|E;D^{+*}D^-,SL\rangle)\nonumber\\ +&\int_{M_{0V0V}}^\infty \mathrm{d}E \sum_{S,L}\frac{f^{0V0V}_{SL}(E)} {z_X-E}|E;D^{0*}\bar
D^{0*},SL\rangle+\int_{M_{+V-V}}^\infty \mathrm{d}E\sum_{S,L}\frac{f^{+V-V}_{SL}(E)} {z_X-E}|E;D^{+*}D^{-*},SL\rangle\Big) ,
\label{eq:X3872-wave-funtion}
\end{align}
where $N_B=\eta'(z_X)^{-1/2}$ is the normalization factor. The continuum states include $D^0\bar{D}^{0*}$,${D}^{0*}\bar
D^0$,  $D^\pm D^{\mp*}$,
$D^{0*}\bar{D}^{0*}$, $D^{+*}D^{-*}$ and the corresponding thresholds
and form factors are denoted using  subscript and superscripts
${00V}$, ${+- V}$, ${0V0V}$, and ${+V-V}$ respectively, with charge
conjugate states sharing the same
quantities. The $D_s^\pm D_s^{*\mp}$ and $D_s^{*\pm}
D_s^{*\mp}$ continuum states are not considered because it is much
more difficult to produce the $s\bar s$ from the vacuum.
The coupling
form factor  represents the interaction between  bare $c\bar c(2^3P_1)$
state from the GI model  and the continuum states.

Since the process from $c\bar c$ to  $J/\psi\rho$ and $J/\psi\omega$
are OZI-suppressed, we do not  include $J/\psi\rho$ and  $J/\psi\omega$
in Eq.~(\ref{eq:generalFriedrichs}). However, the processes from
$D\bar D^*$ and $D^*\bar D^*$ to $J/\psi\rho$ and $J/\psi\omega$  are not
OZI suppressed. Furthermore, the relative ratio of finding
$D\bar{D}^*$ in the X(3872) state is dominant over that of finding
$c\bar{c}$ in our previous study~\cite{Zhou:2017dwj}. Thus, the contribution to the transition amplitude from $X(3872)$ to
$J/\psi\rho$ and $J/\psi\omega$ is expected to come mainly from the continuum
components, and the contribution from the $c\bar c$ component can be
ignored.

Thus,  the next task is to compute the transition amplitude from the
continuum components to final states $|\rho J/\psi\rangle$ and $|\omega
J/\psi\rangle$. This will be the main topic of the next section.

\subsection{The quark rearrangement model}

To evaluate the transition amplitude of the continuum component to the
$J/\psi\omega$ and $J/\psi\rho$ states, a simple approach is to use the
quark rearrangement method developed by Barnes and
Swanson~\cite{Barnes:1991em}. One of the merit of applying the BS
model in our scheme
 is that the interaction in this model contains the one-gluon-exchange
color Coulomb and spin-spin interaction and linear scalar confinement
terms,
which are the nonrelativistic approximation of  the GI model used here in
determining the wave
functions of the bare states.
 The interaction Hamiltonian of the BS
model in the coordinate space is
\bqa
H_I=\sum_{ij}\big\{\frac{\alpha_s}{r_{ij}}-\frac{8\pi\alpha_s}{3m_i m_j}\vec{S}_i \cdot  \vec{S}_j\delta(\vec{r}_{ij})-\frac{3b}{4}r_{ij}\big\}\mathbf{F_i}\cdot\mathbf{F_j}
\eqa
where the sum runs over all pairs $(i,j)$ of quarks and antiquarks in
the hadrons. The $\mathbf{F_i}$ denotes the color generator, and
$\alpha_s$ the running coupling constant,  $b$ the coupling strength
of the linear potential. However, it is more convenient to make
calculations in the momentum representation in accord with the
extended Friedrichs model.

\begin{figure}[h]%
\begin{center}%
\includegraphics[height=30mm]{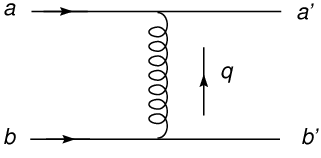}
\caption{\label{fig:TransT} Momentum definitions in the quark-quark transition amplitude.}
\end{center}%
\end{figure}

\begin{figure}[h]%
\begin{center}%
\includegraphics[height=20mm]{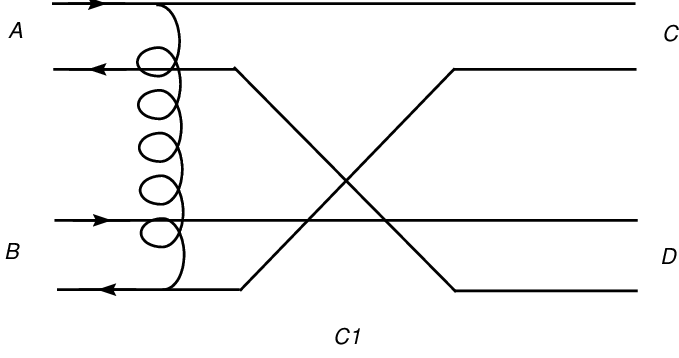}
\includegraphics[height=20mm]{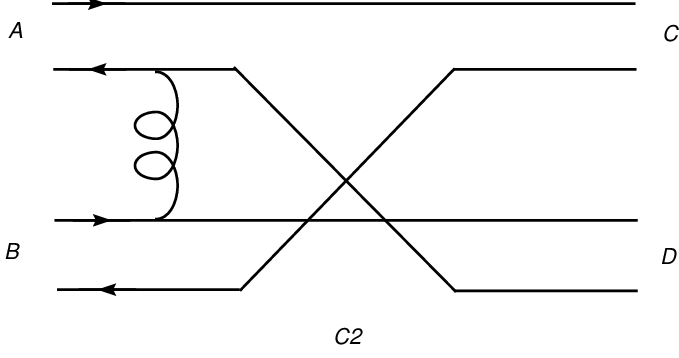}
\includegraphics[height=20mm]{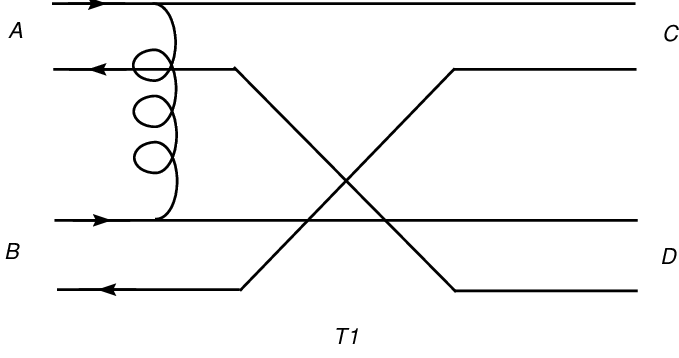}
\includegraphics[height=20mm]{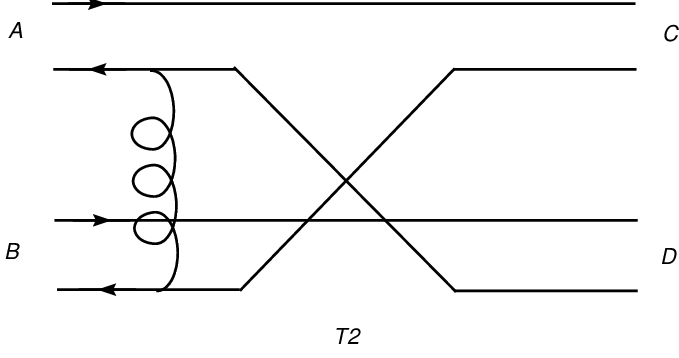}
\caption{\label{rearrange} The four quark rearrangement diagrams of $AB\rightarrow CD$ meson-meson scatterings. The arrows represent the quark line directions.}
\end{center}%
\end{figure}

The main spirit in this model in calculating the  meson-meson
scattering amplitude is to evaluate the lowest order  Born diagrams
of the interchange processes of two constituent
quarks, with the others being considered as spectators. The main
formulas are summarized in the following and the details can be found
in Refs.~\cite{Barnes:1991em,Barnes:1999hs}.
 The momenta for the initial and final constituent quarks are denoted
as $\vec a$, $\vec b$, $\vec{a}'$, and $\vec{b}'$ as shown in
Fig.\ref{fig:TransT} and one can define the linear combinations $\vec q$, $\vec p_1$, and $\vec p_2$ as  $\vec q=\vec a'-\vec a=\vec b-\vec b'$, $\vec p_1=(\vec a+\vec a')/2$, and $\vec p_2=(\vec b+\vec b')/2$.
In the meson-meson transition amplitude of $AB\rightarrow CD$,
conservation of three-momentum implies that the matrix element is of the
form
\bqa
\langle CD|H_I|AB\rangle={ (2\pi)^{-3}}T_{fi}\delta(\vec A+\vec B-\vec C-\vec D).
\eqa
The spin indices of $A$, $B$, $C$, and $D$ mesons are not written out explicitly.
Four kinds of scattering diagrams, ``capture$_1$"$(C_1)$, ``capture$_2$"$(C_2)$, ``transfer$_1$"$(T_1)$, ``transfer$_2$"$(T_1)$
are considered as shown in  Fig.\ref{rearrange}, according to which
pair of the constituents are involved in the interaction~\cite{Barnes:1991em}.
The  $T$-matrix element of every diagram could be represented as the
product of signature, flavor, color, spin, and space factors. The
signature factor is $(-1)$ for all diagrams. The color factor is
$(-4/9)$
for two capture diagram~($C_1$ and $C_2$) and $(4/9)$ for two transfer
diagram~($T_1$ and $T_2$). The flavor factor is obtained by the
overlap of the flavor wave functions of the initial and final states.
The spin factors for $C_1$, $C_2$, $T_1$, and $T_2$ diagrams in
different interaction terms are listed in
Table.~\ref{Table:spinfactor}.

\begin{table}[htp]
\begin{center}
\caption{\label{Table:spinfactor}Compilation of the spin factors for
$C_1$, $C_2$, $T_1$, and $T_2$ diagram in the spin-spin hyperfine, color
Coulomb, and linear potential terms with the total spin of two mesons being 1~\cite{Barnes:1991em}.}
\begin{tabular}{|c|c|c|c|c|c|c|c|c|}
\hline
$(S_A,S_B)\rightarrow (S_C,S_D)$ & \multicolumn{4}{|c|}{$(0,1)\rightarrow (1,1)$} & \multicolumn{4}{|c|}{{$(1,1)\rightarrow (1,1)$}} \\
\hline
  & $C_1$ &$C_2$ & $T_1$&$T_2$& $C_1$ &$C_2$ & $T_1$&$T_2$ \\
\hline
spin-spin & \ \ $\frac{1}{4\sqrt{2}}$ \ \  & $\ \ \frac{1}{4\sqrt{2}}$ \ \  &\  $-\frac{1}{4\sqrt{2}}$ \   & \ $-\frac{1}{4\sqrt{2}}$ \ & \ \ \ \  $0$\ \ \ \  &\ \ \ \   $0$ \ \ \ \   &\ \ \   $-\frac{1}{2}$ \ \ \    & \ \ \    $\frac{1}{2}$ \ \ \   \\
\hline
Coulomb & \multicolumn{4}{|c|}{$\frac{1}{\sqrt{2}}$  } & \multicolumn{4}{|c|}{{$0$}} \\
\hline
linear &\multicolumn{4}{|c|}{$\frac{1}{\sqrt{2}}$  } & \multicolumn{4}{|c|}{{$0$}} \\
\hline
\end{tabular}
\end{center}
\end{table}%

The space overlap factor is obtained by
\bqa
I_{fi}^{space}(AB\rightarrow CD)=\int\int\mathrm{d}^3 q\mathrm{d}^3 p \Phi_C^*(\vec k_C)\Phi_D^*(\vec k_D)T^{pot}_{fi}(\vec q)\Phi_A(\vec k_A)\Phi_B(\vec k_B)
\eqa
where the $\vec k_i=\frac{m_1\vec k_2-m_2\vec k_1}{m_1+m_2}$ denotes
the relative momentum of the quark-antiquark pair in the meson. $\vec
q$ and  $\vec p=\vec p_1$ are the only two independent variables, and the $T^{pot}_{fi}(\vec q)$ refer to the interaction potential due to color Coulomb, spin-spin hyperfine and scalar confinement interaction in the momentum representation as
\bqa
T^{pot}_{fi}(\vec q)=\left\{\begin{array}{ll}
                       4\pi \alpha_s(q)/ q^2\,, & \mathrm{color\ \  Coulomb} \\
                      -8\pi\alpha_s(q)/3m_i m_j\,, & \mathrm{spin-spin\ \  hyperfine} \\
                       6\pi b/ q^4\,, & \mathrm{linear\ \  confinement}
                     \end{array}\right.
\eqa
where the $m_i$ and $m_j$ are the constituent quark masses of the two interacting constituents.
If the wave functions of the mesons  are described as the simple
harmonic oscillator~(SHO) wave functions, the integrations could be
simplified and the analytical transition amplitudes of all four diagrams could be expressed explicitly in terms of confluent hypergeometric functions~\cite{Barnes:1991em,Barnes:1999hs}.
However, the meson wave function of the GI model are represented in a
large number of SHO wave function basis, so the analytical expressions
can hardly be obtained and we can only calculate the integrations
numerically in a practical manner. In calculating the terms of the
linear confinement interaction in the momentum space, the Hadamard
regularization is used to regularize the divergent integrals to obtain
the finite parts.

\subsection{Transition amplitude of X(3872) to $J/\psi\rho$ and $J/\psi\omega$}
The BS model gives the transition amplitudes represented
by the individual spins and the momenta of the initial and final
mesons, and partial wave decomposition must be made to obtain the
matrix elements in term of the angular momentum eigenstates $\langle
J/\psi\rho,S'L'|H_I|E;D^0\bar
D^{0*},SL\rangle$.
The $J/\psi\rho$ and $J/\psi\omega$ two-particle states  are also
decomposed in partial waves with total spin $S'$ and angular momentum
$L'$. So, the transition amplitudes of the $X(3872)$ to $J/\psi\rho$
with particular $S'$ and $L'$ is expressed as
\begin{align}
M_{S'L'}(X(3872)\rightarrow J/\psi\rho)=&\langle J/\psi\rho,S'L'|H_I|X(3872)\rangle\nonumber\\
=&N_B\Big(\langle J/\psi\rho,S'L'|H_I|c\bar c\rangle
\nonumber\\+&\int \mathrm{d}E \sum_{S,L}\frac{f^{00V}_{SL}(E)} {z_X-E}
\Big(\langle J/\psi\rho,S'L'|H_I|E;D^0\bar
D^{0*},SL\rangle
+\langle J/\psi\rho,S'L'|H_I|E;D^{0*}\bar
D^0,SL\rangle\Big)\nonumber\\
+&\int \mathrm{d}E\sum_{S,L}\frac{f^{+-V}_{SL}(E)} {z_X-E}\Big(\langle
J/\psi\rho,S'L'|H_I|E;D^+D^{-*},SL\rangle+\langle
J/\psi\rho,S'L'|H_I|E;D^{+*}D^{-},SL\rangle\Big)
\nonumber\\
+&\int \mathrm{d}E \sum_{S,L}\frac{f^{0V0V}_{SL}(E)} {z_X-E}\langle J/\psi\rho,S'L'|H_I|E;D^{0*}\bar
D^{0*},SL\rangle\nonumber\\ +&\int \mathrm{d}E\sum_{S,L}\frac{f^{+V-V}_{SL}(E)} {z_X-E}\langle J/\psi\rho,S'L'|H_I|E;D^{+*}D^{-*},SL\rangle\Big)
\label{eq:transitionamplitudeJrho}
\end{align}
where the transition of $\langle J/\psi\rho|H_I|c\bar c\rangle$ will
be omitted due to the OZI suppression, as mentioned above. A similar expression could be obtained for the  $J/\psi\omega$ case.

Attention must be paid to the isospin difference of $\rho$ and
$\omega$ whose flavor wave functions are
$\rho=\frac{1}{\sqrt{2}}(u\bar u-d\bar d)$ and
$\omega=\frac{1}{\sqrt{2}}(u\bar u+d\bar d)$ in ideal mixing,
respectively, which lead to opposite signs in the flavor
factors in the terms involving $d$ quarks interchanges in the quark rearrangement model by Barnes and Swanson.

The observed final states of the $X(3872)$ decays are $J/\psi\pi^+\pi^-$ through $\rho$ resonance and $J/\psi\pi^+\pi^-\pi^0$ through $\omega$ resonance. For simplicity, we describe the $\rho$ and $\omega$ resonances by their Breit-Wigner distribution functions~\cite{Zhang:2009bv}, then obtain
\bqa
\Gamma(X\rightarrow J/\psi\rho(\pi^+\pi^-))=\frac{1}{2\pi}\int_{2m_\pi}^{m_X-m_{J/\psi}}\frac{|M(X\rightarrow J/\psi\rho)|^2\Gamma_\rho}{(E-m_\rho)^2+\Gamma_\rho^2/4}\mathrm{d}E,\nonumber\\
\Gamma(X\rightarrow J/\psi\omega(\pi^+\pi^-\pi^0))=\frac{1}{2\pi}\int_{3m_\pi}^{m_X-m_{J/\psi}}\frac{|M(X\rightarrow J/\psi\omega)|^2\Gamma_\omega}{(E-m_\omega)^2+\Gamma_\omega^2/4}\mathrm{d}E,
\label{eq:GammaX}
\eqa
in which the lower limits of the integration are chosen at the experiment cutoffs as in Ref.~\cite{Abe:2005ix,delAmoSanchez:2010jr}.

\section{Numerical results and discussions}

In the numerical calculation, as in \cite{Zhou:2017dwj}, we first
reproduced the meson wave functions in the GI
model~\cite{Godfrey:1985xj} in a large number of SHO basis, and apply them
in the QPC model to obtain the coupling functions in the
Friedrichs-like model. Then, the generalized energy eigenvalues of
Gamow states and their wave functions could be obtained by solving the
Friedrichs-like model. The $X(3872)$ appears as a bound state pole
below the $\bar D^0 D^{0*}$ threshold and its wave function is
obtained. Using the BS model and Eq.(\ref{eq:transitionamplitudeJrho})
 one can then calculate the transition amplitudes
of the $X(3872)$ to the final states,  and the branch ratio is thus obtained.

The parameters used in our calculation include all the parameters in
the GI model, BS model and the $\gamma$ parameter in the QPC model.
All the parameters in GI model are fixed using the original GI's
values in order to obtain the meson wave
functions including the bare $c\bar c$ state, the charmed mesons and $\rho$,
$\omega$.
The same parameters are used in the BS
 without introducing any new parameter since these
two models share similar interaction terms.
The values of the
two-particle thresholds in the Friedrichs-like model are chosen as
their physical masses. Thus, all the parameters except for $\gamma$,
symbolizing the quark pair production rate from the vacuum, are fixed
at the values in the GI model~\cite{Godfrey:1985xj}.
 The $\gamma$ parameter is chosen around $4.0$ to produce a
$X(3872)$ mass in $3.8710\sim 3.8717$
MeV in order to be consistent with the experiment.
 There is also no
cufoff parameter since the higher energy effects are suppressed in the
integration by the form factor and the denominator, and thus the
energies in the integration are integrated up
to infinity. In this calculation, an explicit isospin breaking effect
is caused by considering the mass difference of $u$ and $d$ quarks with
$(m_u+m_d)/2=220\ \mathrm{MeV}$ and $m_d-m_u=5\
\mathrm{MeV}$~\cite{Godfrey:1985xj}.  The isospin breaking effects can
also be introduced into the  production rate parameter through this
mass difference by
$\gamma=\frac{m_u+m_d}{2m_q}\gamma_0$, where $m_q$ is the constituent quark mass of
the produced pair~\cite{Ackleh:1996yt}.

GI's wave functions of the bare meson states are approximately solved
by expanding them in $30$ SHO wave function basis. The wave function
of the $X(3872)$ state as a bound state is also expressed as in Eq.
(\ref{eq:X3872-wave-funtion}).  Then we can evaluate the partial-wave transition
amplitudes of  neutral or charged $D\bar D^*$  to $J/\psi\rho$  or
$J/\psi\omega$. In the BS quark rearrangement model, the spin and
angular momentum are conserved separately, so the partial $L$-wave of
the $J/\psi\rho,\omega$  states can only coupled to the same partial
$L$-wave part in the $D\bar D^*$ continuum term. Since the
vector-vector $D^*\bar D^*$ could only have the $D$-wave continuum in
the QPC coupling, they will only contribute to the $D$-wave
$J/\psi\rho$ or $J/\psi\omega$. The magnitudes of the  $D$-wave scattering
amplitudes of $D^*\bar D^*$ to  $J/\psi\rho$ or
$J/\psi\omega$ are highly suppressed and they are several orders of
magnitude lower
than that of $S$-wave $\bar D D^*$ amplitudes. So, the main
contributions are from the $D\bar D^*$ continuum parts. Furthermore,
the $D^*D^*$ continuum component in $X(3872)$ is very small. The
ratio of
``elementariness" and  ``compositeness" of the different components in the $X(3872)$ is about
$Z_{c\bar c}:X_{\bar D^0 D^{0*}}: X_{ D^+ D^{-*}}: X_{\bar D^* D^*} = 1:(2.67\sim 8.85):(0.45\sim 0.46):0.04$
according to Eq.~(\ref{eq:compositeness}) as $\gamma$ is tuned
to obtain the $X(3872)$ mass in $3.8710\sim 3.8717$
MeV. It is natural that the compositeness is sensitive
to the closeness of the pole to the $\bar D^0 D^{0*}$ threshold due to
the denominator in the integrand in Eq.~(\ref{eq:compositeness}).
This denominator will amplify the form factor near the threshold, and
thus the integral is sensitive to the near-threshold behaviors of coupling
form factor $f_{SL}(E)$. Since the form factor is calculated using the
wave function, different choices of the wave functions may also affect
the behavior of the form factor near the threshold. Here, we choose
the well-accepted GI's wave functions as the input, which is more
accurate than just using a phenomenological monopole form factor like
in \cite{Takizawa:2012hy, Meng:2007cx}.

\begin{figure}[t]%
\begin{center}%
\includegraphics[height=30mm]{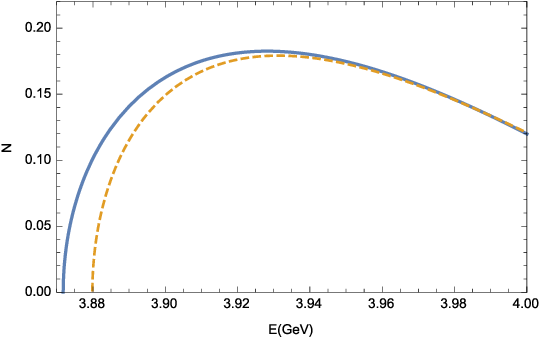}
\includegraphics[height=30mm]{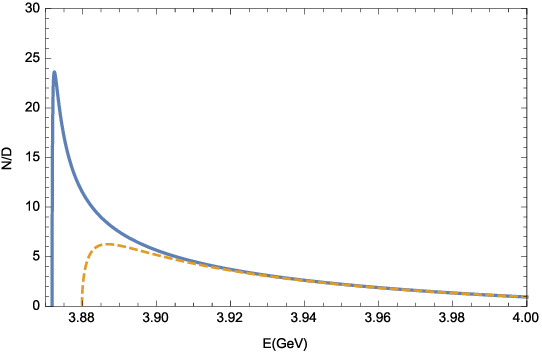}
\caption{\label{fig:compare-neutural-charge} Comparison of the
integrand of Eq.~(\ref{eq:transitionamplitudeJrho}) for charged and
neutral channels  without or with the denominators (N or N/D)
respectively. The solid line represents the neutral $\bar D D^*$
channel and the dashed one the charged channel. The left figure shows
the contributions of the numerator only and the right figure shows the
whole integrand with the denominators.}
\end{center}%
\end{figure}

The isospin breaking effect is caused by the mass
difference of the constituent $u$ and $d$ quark. This leads
to
two consequences in the calculation. First, it causes the mass
difference of the $\bar D^0 D^{*0}$ and $\bar D^+ D^{*-}$.  If the
isospin symmetry is respected, the $\bar D^0 D^{*0}$ and $\bar D^+
D^{*-}$ will be degenerate and their contributions to the scattering
amplitude
to $J/\psi\rho$ from corresponding terms in
Eq.~(\ref{eq:transitionamplitudeJrho}) will be totally canceled but
not in the scattering amplitude to
$J/\psi\omega$. Now that the threshold for $\bar D^0
D^{*0}$ is lower than  $\bar D^+ D^{*-}$ by about $8$ MeV,  $\bar D^0
D^{*0}$ will give a nonzero contribution to the amplitude to $J/\psi
\rho$ below
$\bar D^+ D^{*-}$ threshold. Above $\bar D^+ D^{*-}$ threshold, in general, the nearer it is to the
threshold, the larger  the
difference between the two amplitudes due to the mass difference is.
In addition, the $\gamma$ symbolizing the quark pair
creation strength from the vacuum also depends on the quark masses,  which also leads to a smaller coupling to
the $\bar D^+ D^{-*}$ continuum than to the  $\bar D^0 D^{0*}$ continuum. It is natural that the heavier
components are more difficult to produce. These two effects caused by
the explicit isospin-breaking of $u$ and $d$ quark masses are still
quite tiny since $m_d-m_u$ is about 5 MeV in the GI model. However, it
is greatly amplified by the denominators of $\frac{1}{z_X-E}$ in
Eq.~(\ref{eq:transitionamplitudeJrho}) because the mass of $X(3872)$
is very close to the $\bar D^0 D^{*0}$ threshold. To demonstrate this
mechanism clearly, we plot the numerator part in the
integrand and the
whole integrand of
neutral and charged $\bar D D^*$ terms in
Eq.~(\ref{eq:transitionamplitudeJrho}), respectively, for the
amplitude to $J/\psi\rho$ in
Fig.~\ref{fig:compare-neutural-charge} for comparison. The left figure
shows the absolute values of the numerators of neutral~(solid) and
charged~(dashed) $\bar D D^*$ contributions in
Eq.~(\ref{eq:transitionamplitudeJrho}), which are slightly different.
However, the whole integrand with the $\frac{1}{z_X-E}$
denominator shown in the right figure are totally different.
We see that near the threshold the differences between the two integrand are
greatly enhanced by the denominator and hence give a sizable
contribution the the amplitude.

After the amplitudes are calculated by integrating up to infinity in (\ref{eq:transitionamplitudeJrho}), one could obtain the relative ratio of
the transition amplitudes, $\frac{M(X\rightarrow
J/\psi\omega)/\sqrt{P_{J/\psi\omega}}}{M(X\rightarrow
J/\psi\rho)/\sqrt{P_{J/\psi\rho}}}=3.24\sim 2.43$, with $X(3872)$
 mass ranging from 3871.0 MeV to 3871.7 MeV, which is
consistent with the expectation of Suzuki\cite{Suzuki:2005ha} to
produce a reasonable isospin breaking.
In principle, to obtain the decay widths using Eq. (\ref{eq:GammaX}), the $\rho,\omega$
masses in the scattering amplitudes of $\bar DD^* \rightarrow
J/\psi\rho,\omega$ should be regarded as variables. However, since the
numerical calculation of the scattering amplitudes involve
six-dimension integration, the integration would be difficult and
costs too much
computer time. We have checked that the numerical values of the
scattering amplitudes changes very slowly when the masses of
$\rho,\omega$ changes. Thus, in a practical manner, we did not calculate the
integration directly but approximately calculated the scattering
amplitudes with the $\rho,\omega$ masses fixed at their central values
in PDG and product the phase space kinematic factors.
If we cut off  the $\pi^+\pi^-$  invariant mass at $0.45$ GeV
and the $\pi^+\pi^-\pi^0$ invariant mass at 0.75 GeV as chosen by
Belle and \textsl{BABAR}~\cite{Abe:2005ix,delAmoSanchez:2010jr}, the
kinematic space ratio will be
\bqa
\frac{\int_{3m_\pi}^{m_X-m_{J/\psi}}\frac{|P_{J/\psi\omega}|\Gamma_\omega}{(E-m_\omega)^2
+\Gamma_\omega^2/4}\mathrm{d}E}{\int_{2m_\pi}^{m_X-m_{J/\psi}}\frac{|P_{J/\psi\rho}|\Gamma_\rho}
{(E-m_\rho)^2+\Gamma_\rho^2/4}\mathrm{d}E}=0.088\sim 0.098
\eqa
as the mass of $X(3872)$ ranges from 3871.0 MeV to 3871.7 MeV
and the branch fraction will be
\bqa
\frac{\Gamma_{J/\psi\pi^+\pi^-\pi^0}}{\Gamma_{J/\psi\pi^+\pi^-}}=(3.24\sim
2.43)^2 \times (0.088\sim 0.098)=0.92\sim 0.58.
\eqa
This numerical results are consistent with the measured ratio
$1.0\pm 0.4\pm 0.3$
by Belle~\cite{Abe:2005ix} and
$0.8\pm 0.3$
 by
\textsl{BABAR}~\cite{delAmoSanchez:2010jr}.
If the lower limits are chosen at $2m_\pi$ and $3m_\pi$ physical masses respectively, the ratio of kinematic factors will be $0.161\sim 0.169$ and
\bqa
\frac{\Gamma_{J/\psi\pi^+\pi^-\pi^0}}{\Gamma_{J/\psi\pi^+\pi^-}}=(3.24\sim 2.43)^2 \times (0.161\sim 0.169)=1.69\sim 0.99.
\eqa

\begin{figure}[t]%
\begin{center}%
\includegraphics[height=30mm]{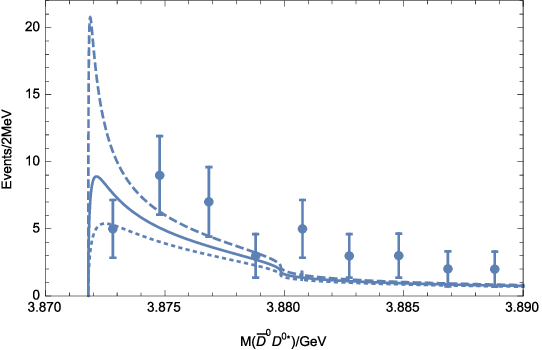}
\includegraphics[height=30mm]{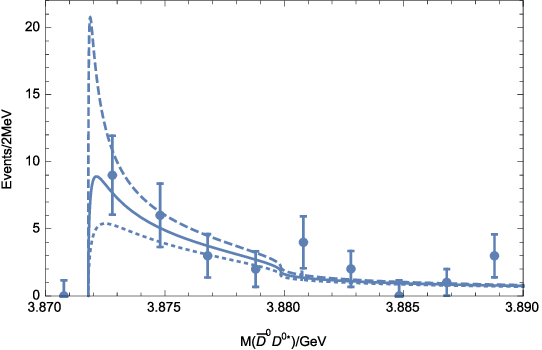}
\includegraphics[height=30mm]{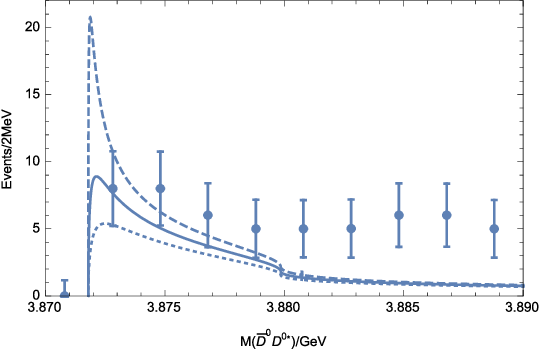}
\caption{\label{fig:DDstardistribution} The experimental $\bar D^0
D^{0*}$ mass distributions in $B\rightarrow X(\bar D^0 D^{0*})K$
decays of \textsl{BABAR}~(left)~\cite{Adachi:2008sua} and
Belle~(middle and right)~\cite{Aubert:2007rva} compared with $r|T(\bar
D^0 D^{0*}\rightarrow \bar D^0 D^{0*})|^2\cdot phasespace$, where $r$
is a rescaling factor chosen by hand. The dotted, solid, and dashed
lines represent the curves with $X(3872)$ mass at 3.8710, 3.8714, and
3.8717 GeV, respectively.}
\end{center}%
\end{figure}

As a byproduct, this scheme can also provide a qualitative interpretation to
the $\bar D^0 D^{0*}$ mass distribution in
$B\rightarrow \bar D^0 D^{0*}K$
processes by Belle and
\textsl{BABAR}~\cite{Aubert:2007rva,Adachi:2008sua}. Since the
multi-channel scattering
amplitudes  could be obtained similar to
Eq.~(\ref{scatteringSmatrix}) as in Ref.~\cite{Xiao:2016mon}, the
$T$-matrix of $\bar D^0 D^{0*}$ scatterings could be obtained as
\bqa\label{scatteringTmatrix}
T_{SL}(\bar D^0D^{0*}\rightarrow \bar D^0D^{0*})= \frac{| f^{\bar D^0D^{0*}}_{SL}(E)|^2}{\eta^+(E)}.
\eqa
If the weak interaction vertex in $B\rightarrow X(\bar D^0
D^{0*})K$ decays is supposed to be a smooth mildly changing factor and
could be simulated by a constant, the
mass distribution of the $\bar D^0 D^{0*}$ final states is
proportional to the $|T(\bar D^0 D^{0*}\rightarrow \bar D^0
D^{0*})|^2$ and the phase space factors. A qualitative agreement could
be found between the calculations and the experiment data up to a
rescaling factor, as shown in Fig.~\ref{fig:DDstardistribution}. The
dotted, solid, and dashed lines in Fig.~\ref{fig:DDstardistribution}
represent the $X(3872)$ mass at 3.8710, 3.8714, and 3.8717 GeV,
respectively, up to a rescaling factor.

\section{Summary}

In this paper,  we perform a calculation on the branching ratio of
$X(3872)$ transition to $J/\psi\pi^+\pi^-\pi^0$ and $J/\psi\pi^+\pi^-$
based on the Friedrichs-like scheme combined with  the quark
rearrangement model and QPC model. In our previous
work~\cite{Zhou:2017dwj}, the first
excited $P$-wave charmonium state spectrum is reproduced using the same
Friedrichs scheme combined with QPC with GI's bare spectrum as input,
and the results are consistent with the
experiment, which demonstrates the reasonability of this scheme. In
fact, this scheme provides a general way to incorporate the hadron
interactions into the GI quark model. The $X(3872)$ is dynamically generated
as a bound state just below the $D^{*0}\bar D^0$, composed of a dominant $\bar D^0
D^{0*}$ continuum component of about  $64\sim 85\%$, a $c\bar c$
component of about $10\sim 24\%$, and other continuum.
The $X(3872)$ wave function can  be explicitly expressed. Based on these
information, in present paper we studied the isospin
breaking effect within this framework. Since the $X(3872)$  is mostly composed
of continuum and the $c\bar c$ contribution is also suppressed by OZI
rule, we can only consider the continuum contributions to the
decay amplitude. In the spirit of the quark rearrangement model, by considering the
spin-spin, color Coulomb, and linear potential interactions among the
quarks in different mesons, one can obtain the transition amplitudes
for $X(3872)$ to $J/\psi\omega$ and $J/\psi\rho$, using the
wave function obtained from the Friedrichs-scheme. By taking
into account the mass difference of the final states, we obtain the numerical
result
$\frac{\Gamma_{J/\psi\pi^+\pi^-\pi^0}}{\Gamma_{J/\psi\pi^+\pi^-}}=0.92\sim
0.58 $ as the $X(3872)$ mass changes from 3871.0 MeV to 3871.7 MeV,
which is comparable to the experiments.  We notice that to provide a
reasonable magnitude of the isospin breaking amplitude, the proximity
of the $X(3872)$ position to the $D^0D^{0*}$ plays an important role. It greatly
amplifies the amplitude to $D^0\bar D^{0*}$ near threshold which
causes a large isospin breaking effect in the amplitude as shown in
Figure \ref{fig:compare-neutural-charge}. This amplification effect
also presents in the integration in the compositeness calculation.
Thus,
the precise near-threshold behavior of the form factor is important for this
result to be solid. Since our calculations are based on
 GI's wave functions, which supposedly are the
{\it de facto} standard in the literature, they presumably
provide a more precise form factor which makes our results more
solid and convincing.

In this calculation, because the interaction
Hamiltonian between the quarks of different meson in the quark
rearrangement model is similar to GI's quark potential model,
there is no new parameters introduced in the calculation.
All the model parameters except the quark
production rate from the vacuum, $\gamma$, are fixed at the
well-accepted GI model. The $\gamma$ parameter is also chosen such that the $X(3872)$ is around the experimental mass
and all the other observed first excited $P$-wave states are
reproduced well as in \cite{Zhou:2017dwj}. One of the merits of using our scheme is that the
high energy contribution in the integration in calculating the decay
amplitudes is naturally suppressed by the form factor obtained from
QPC model based on the wave function from GI's relativized quark
model. Thus, unlike in \cite{Suzuki:2005ha} and \cite{Meng:2007cx}
where the result is cutoff dependent, here no cutoff is introduced and
thus the result is more robust.

\begin{acknowledgments}
Helpful discussions with Dian-Yong Chen, Hai-Qing Zhou, Ce Meng, and Xiao-Hai Liu are appreciated.
Z.X. is supported by China National Natural
Science Foundation under contract No.  11105138, 11575177 and 11235010. Z.Z is supported by the Natural Science Foundation of Jiangsu Province of China under contract No. BK20171349.
\end{acknowledgments}

\bibliographystyle{apsrev4-1}
\bibliography{Ref}

\end{document}